\documentclass[aps,twocolumn,amsmath,amssymb,nofootinbib,superscriptaddress]{revtex4-1}

\usepackage{graphicx}

\usepackage{amsmath}
\usepackage{amsfonts}
\usepackage{amssymb}
\usepackage{amsthm}
\usepackage{mathtools}
\usepackage{color}

\usepackage{hyperref}

\newcommand{\beq}{\begin{equation}}
\newcommand{\eeq}{\end{equation}}

\begin{document}

\author{Christian Arenz} 
\affiliation{Frick Laboratory, Princeton University, Princeton NJ 08544, US}

\author{Herschel Rabitz} 
\affiliation{Frick Laboratory, Princeton University, Princeton NJ 08544, US}

\title{Controlling Qubit Networks in Polynomial Time}

\date{\today}

\begin{abstract}
Future quantum devices often rely on favourable scaling with respect to the number of system components. To achieve desirable scaling, it is therefore crucial to implement unitary transformations in a time that scales at most polynomial in the number of qubits. We develop an upper bound for the minimum time required to implement a unitary transformation on a generic qubit network in which each of the qubits is subject to local time dependent controls. Based on the developed upper bound the set of gates is characterized that can be implemented polynomially in time. Furthermore, we show how qubit systems can be concatenated through controllable two body interactions, making it possible to implement the gate set efficiently on the combined system. Finally a system is identified for which the gate set can be implemented with fewer controls. The considered model is particularly important, since it describes electron-nuclear spin interactions in NV centers.  
\end{abstract}

\maketitle
Achieving accurate control and scalability lie at the heart of every functioning quantum information processing device. Thus, a vital goal is to design algorithms that can be implemented efficiently. In particular, in the gate model of quantum information processing an efficient algorithm should scale polynomially in the number of gates used to carry out the computation. Through a \emph{universal gate set} every algorithm described by a unitary transformation can be implemented up to some degree of accuracy. However, a simple counting argument shows that most of the unitary transformations cannot be implemented efficiently \cite{NielseChuang}. Quantum control theory allows for  implementing the final unitary transformation directly through optimized classical control fields \cite{BookDalessandro, ControlRev1, ControlRev2}. This has the advantage that, if the procedure can be done efficiently, there is no need for constructing gate sequences. Instead, optimization algorithms such as a gradient based search \cite{Grape, SpinQubits}, learning control \cite{Hersch1, Daniel}, or genetic algorithms \cite{Hersch2, Manu}, may be used to pre-calculate or learn the classical control fields that implement the desired unitary transformation. In fact, it has been shown that the complexity of both approaches, i.e., calculating control pulses and designing gate sequences is the same \cite{Nielsen1, Nielsen2}. 

Similar to a universal gate set, for a \emph{fully controllable} system every unitary transformation contained in the special unitary group $\text{SU}(2^{n})$ is reachable through switchable controls. In order to implement a goal unitary gate $U_{g}$ efficiently, it is crucial that the length of the control pulses, henceforth referred to as the \emph{minimum gate time} $T$, scales at most polynomially in the number of qubits. Across the quantum information sciences, a reasonable scaling of the minimum gate time and the charaterization of the efficiently implementable gates is of particular importance in order to determine whether a system is suitable for quantum information tasks. Unfortunately, the determination of the minimum gate time has remained a major technical challenge to overcome for moving the field towards practical applications. In this letter we make a significant step towards solving this problem by developing an upper bound for the minimum gate time under the assumption that sufficient control recources are available. As illustrated in figure 1, this allows for determining the set of gates that \emph{provably} can be implemented efficiently. 

Although  substantial progress has recently been made by characterizing graphs that can be controlled efficiently \cite{Lloyd}, the characterization of the set of gates that can be reached in polynomial time and the corresponding number of controls required is still unknown. Moreover, it remains challenging to identify physical models that obey the criteria developed in \cite{Lloyd}. 
  \begin{figure}[!h]
  \includegraphics[width=1.0\columnwidth]{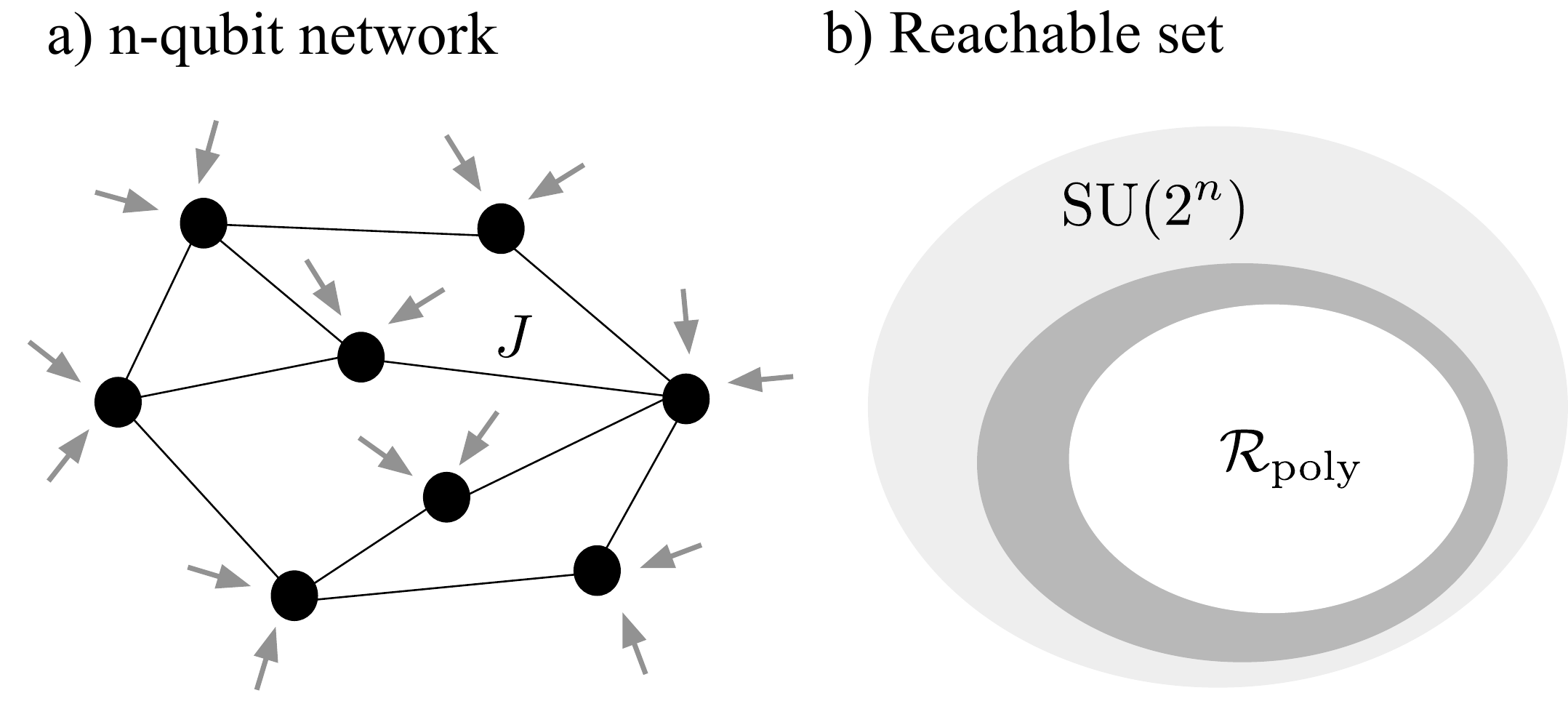}
 \caption{\label{fig:illustration} Illustration of one of the paper's main results: a) for a generic qubit network \eqref{eq:driftH} in which each of the qubits is subject to two local controls \eqref{eq:controlH} (grey arrows), b) the set of gates $\mathcal R_{\text{poly}}$ (white area) that (provably) can be implemented in a time that scales at most polynomialy in the number of qubits is characterized (see \eqref{eq:setpolynomial}). The dark grey area represents the set of gates that can be reached with time optimal methods in polynomial time.}
\end{figure}

The main quantitative result of this letter is the development of the upper bound 
\begin{align}
\label{eq:boundfinal}
T(\bold{a})\leq\frac{l(\bold{a})}{J}\left(\Vert \bold{a}\Vert_{\infty}+\frac{\pi l(\bold{a})(l(\bold{a})-1)(n-2)\Vert \bold{a}\Vert_{\infty}^{2}}{2\sqrt{2}\epsilon}   \right),	
\end{align}	
for the minimum gate time to implement a goal unitary transformation $U_{g}(\bold{a})$ up to some error $\epsilon$ for a generic $n$-qubit graph \eqref{eq:driftH} in which each of the qubits is subject to two local controls \eqref{eq:controlH}. As illustrated in figure \ref{fig:illustration} a), the qubits (black circles) interact via two body interactions (solid lines) where $J$ is the smallest coupling constant present in the graph. The generator of the goal unitary transformation is characterized by $l(\bold{a})$ real parameters summarized in the vector $\bold{a}$ with $\Vert\bold{a}\Vert_{\infty}$ being the vector infinity norm, i.e. the largest parameter. 
  
\begin{figure}[!h]
  \includegraphics[width=1.0\columnwidth]{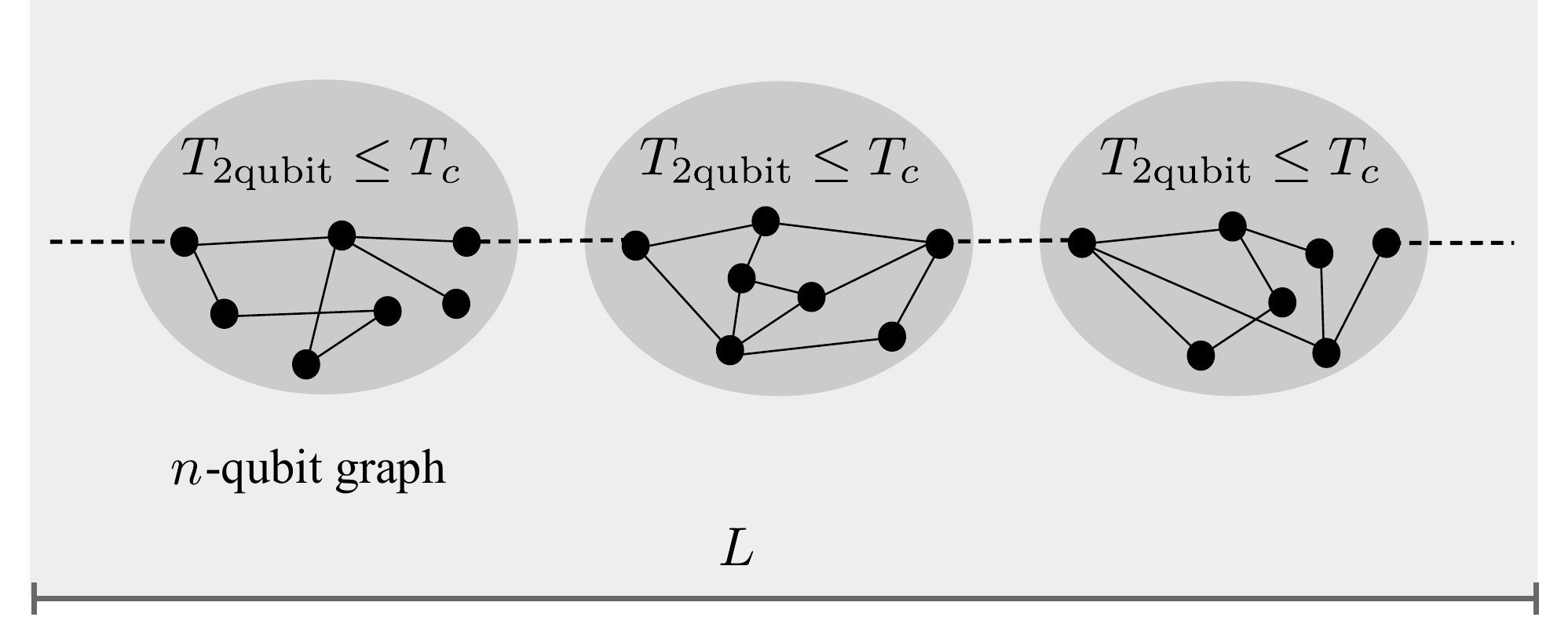}
 \caption{\label{fig:concatenation} Illustration of how $L$ qubit graphs, each consisting of $n$ qubits, can efficiently be concatenated through controllable two body interactions (dotted lines). Assuming that on each qubit graph any two qubit gate can be implemented in a time smaller than $T_{c}$, for  $\Vert\bold{a}\Vert_{\infty}^{2}=\mathcal O(1)$ and $l(\bold{a})=\mathcal O(n)$, we then have for the total system $T(\bold{a})\leq \mathcal O(T_{c}(Ln)^{5})$.}
\end{figure}

  One way to obtain an upper bound on $T$ is to find a specific way to implement a generic unitary transformation and upper bound the corresponding time. The procedure that is used here can be summarized by the following steps, with details found below and in the online material \cite{OnlineMaterial}:
\begin{enumerate}
\item Due to the assumption that each qubit is subject to two unconstrained orthogonal controls, a decoupling sequence allows to select arbitrary two body interactions instantaneously (see Eq. \eqref{eq:twobodyunitary}).
\item A sequence formed by such two qubit unitaries (see Eq. \eqref{eq:sequence}) allows for creating unitary operations which are generated by $k$-body interaction terms in a time that scales linearly in $k$ (see Eq. \eqref{eq:taB}). 
\item Finally, unitary transformations that are generated by linear combinations of $l$, $k$-body interaction terms can be created (up to an error $\epsilon$)  using a Trotter sequence (see Eq. \eqref{eq:boundF}). 
\end{enumerate}
As illustrated in figure \ref{fig:illustration} b), the bound \eqref{eq:boundfinal} allows to conclude that the gate set 
\begin{align}
\label{eq:setpolynomial}
\mathcal R_{\text{poly}}=\{U(\bold{a})\in \text{SU}(2^{n})~|~l(\bold{a}),~\Vert\bold{a}\Vert_{\infty} \leq \mathcal O(\text{poly}(n))\},
\end{align} 
can be implemented on a qubit graph in which each qubit is subject to two local controls in a time that scales at most polynomially in the number of qubits $n$, and, moreover, enables for characterizing the Hamiltonians that can be simulated efficiently. We furthermore show that for a specific system the gate set $\mathcal R_{\text{poly}}$ can be implemented with less controls. Moreover, a strategy is presented for efficiently scaling the system by controlling two body interactions (schematically represented in figure \ref{fig:concatenation}), thereby paving the way towards functioning quantum devices.     

In order to derive the bound \eqref{eq:boundfinal}, we first introduce some mathematical terminology. We remark that our findings are a proof of feasibility rather than a strategy to implement gates in a time optimal manner, which remains a practical challenge. A quantum control problem can be expressed as follows. The system of interest is described by a time dependent Hamiltonian of the form 
$H(t)=H_{0}+H_{c}(t)$,	
 where $H_{0}$ is referred to as the drift Hamiltonian and the controls enter in $H_{c}(t)$ via time dependent functions. The aim of quantum control is then to steer the system towards a desired target by shaping the control functions. Here we are interested in implementing a generic target unitary transformation $U_{g}$ on a $n$ qubit system. The first question to consider is whether every $U_{g}$ can be reached, i.e., whether the system is fully controllable. When control enters in a bilinear way in $H_{c}(t)$ \cite{Elliot}, known as the Lie rank criterion \cite{BookDalessandro}, the system is fully controllable iff the controls and drift generate the full algebra (see e.g., \cite{Carenzspinstar, LieAlExample1, LieAlExample2, SimilarModelC1, SimilarModelC2, Symmetries1, Symmetries2} and references therein for examples). More formally, if the system is fully controllable there exist controls which allow implementing every $U_{g}=\exp(\Theta)$ with $\Theta\in \mathfrak{su}(2^{n})$ up to arbitrarily high precision in finite time. Throughout this work the special unitary algebra $\mathfrak{su}(2^{n})$ is expressed in terms of the Pauli operator basis $\{B_{i}\}_{i=1}^{2^{2n}-1}$, in which each $B_{i}$ corresponds to a string of Pauli operators. Every $\Theta\in \mathfrak{su}(2^{n})$ can be written as $\Theta(\bold{a})=\sum_{i=1}^{l(\bold{a})}a_{i}B_{i}$, where the real coefficients are summarized in the vector $\bold{a}$ and we denote by $l(\bold{a})\leq 2^{2n}-1$ the number of its non-zero elements. Except for low dimensional systems \cite{OptimalControlSpeedLimit2, ExactCalc1, ExactCalc2, ExactCalc3, ExactCalc4, Nori}, the minimum gate time $T(\bold{a})$ needed to implement $U_{g}(\bold{a})$ up to some accuracy is not known.  
  
Consider a connected graph $G(V,E)$  where the vertices $V$ and edges $E$ represent qubits and two body interactions, respectively. The most general form of such an $n$-qubit graph is described by the drift Hamiltonian
\begin{align}
\label{eq:driftH}
H_{0}=\sum_{\mathclap{\substack{i\in V, \\  \alpha\in\{x,y,z\}}}}\omega_{\alpha}^{(i)}\sigma_{\alpha}^{(i)}+\sum_{\mathclap{\substack{(i, j)\in E,\\\alpha,\beta\in\{x,y,z\}}}}g_{\alpha,\beta}^{(i,j)}\sigma_{\alpha}^{(i)}\sigma_{\beta}^{(j)},
\end{align}
where $\omega_{\alpha}^{(i)},~g_{\alpha,\beta}^{(i,j)}$ are energy splittings and coupling constants, respectively.  
Here the notation refers to $\sigma_{\alpha}^{(j)}\equiv\openone\otimes \sigma_{\alpha}\otimes  \openone$ where $\sigma_{\alpha}$ with $\alpha\in\{x,y,z\}$ are Pauli spin operators. That is, $\sigma_{\alpha}^{(i)}$ acts only non-trivially on the $i$th qubit. We assume that each qubit is subject to two local controls $\{\sigma_{x}^{(i)},~\sigma_{y}^{(i)}\}$ such that
\begin{align}
\label{eq:controlH}
H_{c}(t)=\sum_{i\in V}(f_{i}(t)\sigma_{x}^{(i)}+h_{i}(t)\sigma_{y}^{(i)}),	
\end{align}
where $f_{i}(t),h_{i}(t)$ are the corresponding control fields which are assumed to be unconstrained. This is a typical assumption in the context of quantum control theory and dynamical decoupling and its crucial for the development of the upper bound below (see the note \cite{Comment1}). Before relaxing the assumption of two orthogonal controls on each qubit, we first describe in more detail how the upper bound on $T(\bold{a})$ can be derived for this control system. Further details of the derivation can be found in \cite{OnlineMaterial}.  

The analysis starts with the form of $H_{c}(t)$, allowing for having two orthogonal controls on each qubit, such that every single qubit gate can be implemented instantaneously \cite{ExactCalc1, ExactCalc2}; moreover, the system is fully controllable \cite{Symmetries1}. Using a decoupling sequence \cite{LVioal1, Dec2} formed by the controls, permits instantaneously selecting arbitrary two body interaction terms \cite{DecLloyd}. Thus, we can implement every unitary transformation 
\begin{align}
\label{eq:twobodyunitary}
U_{\alpha,\beta}^{(i,j)}(k)=e^{\pm ik \sigma_{\alpha}^{(i)}\sigma_{\beta}^{(j)}},~~~~k\in \mathbb R_{+},~~~~\alpha,\beta\in\{x,y,z\},	
\end{align}
 in a time $t=k/g_{\alpha,\beta}^{(i,j)}$ \cite{OnlineMaterial}, noting that each Pauli operator can be rotated intanteously to a generic Pauli operator using local operations. The following analysis makes use of fact that every basis operator $B_{i}$ can be created by a nested commutator of the form $[\cdots,[S_{1},[S_{2},S_{3}]]]$ where $S_{k}\in\mathcal S=\{i\sigma_{\alpha}^{(i)}\sigma_{\beta}^{(j)}\}$, which are referred to as a \emph{generating set} and we refer to the length of the nested commutator as the \emph{depth} $D$ with $[S_{1},S_{2}]$ being a commutator of $D=1$.   Using a sequence of the form 
  \begin{align}
 \label{eq:sequence}
 U_{x,z}^{(2,3)\dagger}(\pi/4)U_{z,y}^{(1,2)}(k)U_{x,z}^{(2,3)}(\pi/4)=\exp(ik\sigma_{z}^{(1)}\sigma_{z}^{(2)}\sigma_{z}^{(3)}),
 \end{align}
 and introducing the smallest coupling constant $J=\min_{i,j,\alpha,\beta}\{g_{\alpha,\beta}^{(i,j)}\}$ present in $H_{0}$, an upper bound for the time $\Delta t$ to create a unitary operation generated by a commutator of depth 1, in \eqref{eq:sequence} a 3-body interaction term, can be found, i.e., $\Delta t\leq \frac{\pi}{2J}$ \cite{OnlineMaterial}.
 
  We remark here that there are other sequences that allow for increasing or decreasing the length of a Pauli string \cite{AnSequence}. Due to the form of the construction \eqref{eq:sequence}, a unitary operation generated by a nested commutator of depth $D$ will then take at most time $D\Delta t$. Thus, the time $\tau(a_{i}B_{i})$ to implement a unitary operation $U_{g}=\exp(a_{i}B_{i})$ is upper bounded by 
  \begin{align}
\label{eq:taB}
\tau(a_{i}B_{i})\leq \frac{1}{J}\left(D(B_{i})\frac{\pi}{2}+|a_{i}|\right),   
\end{align}
which is compared with known results in the online material \cite{OnlineMaterial}. 
Through a Trotter-Suzuki sequence \cite{Trotter} we can further upper bound the time it takes to generate a unitary operator generated by linear combinations of the basis operators up to an error $\epsilon$. We find 
  \begin{align}
  \label{eq:boundF}
  	T(\bold{a})\leq \frac{1}{J}\left(\Vert\bold{a}\Vert_{1}+\frac{\pi K(\bold{a})\sum_{i=1}^{l(\bold{a})}D(B_{i})}{4\sqrt{2}\epsilon}  \right),
  \end{align}
with $\Vert\cdot\Vert_{1}$ being the vector-1 norm, $K(\bold{a})=\frac{1}{\sqrt{2^{n}}}\sum_{j>k}|a_{j}a_{k}|\Vert[B_{j},B_{k}]\Vert$, and $\Vert\cdot\Vert$ is the Hilbert-Schmidt norm. The  scaling in $\epsilon$, explicitly given in \cite{OnlineMaterial}, can be traced back to the use of the Suzuki-Trotter series, and the scaling can be improved using more sophisticated sequences \cite{SopSequence}. An algorithm finding the ``shortest'' path, possibly weighted by the coupling constants, to create a $B_{i}$ would produce the tightest bound.  However, it takes a nested commutator of depth $(n-2)$ to create a basis operator that contains $n$ Pauli operators $\sigma_{\alpha}^{(1)}\sigma_{\beta}^{(2)}\cdots\sigma_{\delta}^{(n)}$. From this operator it takes another $(n-2)$ commutators to create any $B_{i}$. For an illustration we refer to the Lie tree diagram in \cite{OnlineMaterial}. Thus, the depth is upper bounded by $D(B_{i})\leq 2(n-2)$, yielding the bound \eqref{eq:boundfinal}. Provided that $\Vert\bold{a}\Vert_{\infty}$ scales at most polynomially in the number of qubits, we then have as a sufficient criterion for efficiently implementing a goal unitary $U_{g}$ the following result. For the control system \eqref{eq:driftH} and \eqref{eq:controlH}, a unitary gate $U_{g}(\bold{a})$ that is parameterized through $l(\bold{a})$ parameters can be implemented in a time that is at most polynomial in the number of qubits $n$ if $l(\bold{a})\leq \mathcal O(\text{poly}(n))$. Thus, for the control system in  \eqref{eq:driftH} and \eqref{eq:controlH} the set of gates $\mathcal R_{\text{poly}}$ given by \eqref{eq:setpolynomial} can be reached in a time that scales at most polynomially in the number of qubits. In particular for $\Vert\bold{a}\Vert_{\infty}=\mathcal O(1)$ and $l(\bold{a})=\mathcal O(n)$ every $U_{g}$ can be implemented in a time at most of the order $\mathcal O(n^{4})$. However, in general for $l(\bold{a})=2^{2n}-1$ the upper bound scales exponentially $T(\bold{a})\leq \mathcal O(n2^{6n})$. The bound \eqref{eq:boundF} can be directly applied to efficiently simulating the dynamics with Hamiltonians \cite{HamSim1, HamSim2, HamSim3}. For the control system expressed in Eq. \eqref{eq:driftH} and Eq. \eqref{eq:controlH} every Hamiltonian $H=-i\Theta(\bold{a})$ consisting of $l(\bold{a})\leq \mathcal O(\text{poly(n)})$ k-body interaction terms can be simulated efficiently. Since the strategy to obtain \eqref{eq:boundF} is not necessarily time optimal, the actual set of gates that can be reached in polynomial time may be larger.  It would be interesting to see how much the set can be increased using time optimal control methods \cite{TimeOptimalC}. However, the set $\mathcal R_{\text{poly}}$ can certainly be increased  by considering the full expression in \eqref{eq:boundF}. Moreover, one can easily determine the maximum time needed to implement $U_{g}(\bold{a})=\exp(\Theta(\bold{a}))$ by expanding $\Theta$ in the Pauli operator basis and calculating \eqref{eq:boundF}.

 For $l(\bold{a})=1$ the target unitary operation is given by $U_{g}=\exp(a_{i}B_{i})$ and it follows from \eqref{eq:taB} that the time to implement such an operation is upper bounded by $\tau(a_{i}B_{i})\leq \frac{1}{J}(\pi(n-2)+|a_{i}|)$. For instance, every two qubit gate corresponding to a basis operator with two Pauli operators can be implemented in a time that scales at most linearly in the number of qubits. Moreover, gates corresponding to basis operators with $n$ Pauli operators, i.e., $n$ body interaction terms of the form $\sigma_{\alpha}^{(1)}\sigma_{\beta}^{(2)}\cdots\sigma_{\delta}^{(n)}$, can be implemented in linear time as well.
   The bound can be tightened by introducing the geodesic path distance $d(i,j)$ between two qubits $i$ and $j$ as the smallest number of edges in a path connecting the two considered qubits. For example, it follows that the time to create a CNOT gate between qubit $i$ and $j$ is upper bounded $T_{\text{CNOT}}\leq \pi\left(\frac{d(i,j)-1}{J}+\frac{1}{4J}\right)$. Since every two qubit gate can be implemented with at most three CNOT gates \cite{3CNOT}, up to local unitary rotations, we have
    $T_{\text{2qubit}}\leq 3\pi\left(\frac{ d(i,j)-1}{J}+\frac{1}{4J}	\right)$.
     We remark here that this bound is tighter than the bound that would be obtained by simply implementing a CNOT gate on two nearest neighbor qubits followed by SWAP operations \cite{ExactCalc1, ExactCalc2}. The upper bound for $T_{\text{2qubit}}$ describes how much time is maximally needed in order to implement a generic two qubit gate on a qubit graph \eqref{eq:driftH}, provided each qubit can be instantaneously controlled locally. Therefore, the bound for $T_{\text{2qubit}}$ characterizes the time scale for entangling two qubits in a generic qubit network. 
   
   The characterization of the set of gates that can be reached in polynomial time \eqref{eq:setpolynomial} relied on the assumption that each qubit is subject to two orthogonal controls. A natural question is whether the number of controls can be reduced while still being able to implement $\mathcal R_{\text{poly}}$ in a time that scales at most polynomially in the number of qubits. Before presenting an $n$-qubit graph for which this is the case with only $n+1$ controls, we address the question regarding how qubit systems can be concatenated in order to implement $\mathcal R_{\text{poly}}$ on the total system.

\emph{Concatenating systems --}
Suppose we have two $n$-qubit graphs $G_{1}(V_{1},E_{1})$ and $G_{2}(V_{2},E_{2})$ for which the time to implement a generic two qubit unitary on each of the graphs is upper bounded by $T_{c}$. Now, as represented in figure \ref{fig:concatenation}, connect the two graphs with a single controllable two body interaction, say $\sigma_{z}^{(i)}\sigma_{z}^{(j)}$ with $i\in V_{1}$ and $j\in V_{2}$. Importantly, $\{i\sigma_{\alpha}^{(i)}\sigma_{\beta}^{(i^{\prime})},~i\sigma_{\gamma}^{(j)}\sigma_{\delta}^{(j^{\prime})},~i\sigma_{z}^{(i)}\sigma_{z}^{(j)}\}$ with $i,i^{\prime}\in V_{1}$ and $j,j^{\prime}\in V_{2}$ forms a generating set $\mathcal S$. 

Thus, every basis operator $B_{i}\in\mathfrak{su}(2^{2n})$ for the total system can be created through a nested commutator formed by the elements of $\mathcal S$. The time to create a unitary $U_{g}=\exp(a_{i}S_{i})$ with $S_{i}\in \mathcal S$ is upper bounded by $T_{c}$. Therefore it takes $\Delta t\leq 2T_{c}$ to produce a nested commutator of depth $1$ and the depth for a generic basis operator $D(B_{i})$ is upper bounded by $2(2n-1)$. Consequently the time to implement a unitary transformation $U_{g}=\exp(a_{i}B_{i})$ on the total system is upper bounded by 
$\tau(a_{i}B_{i})\leq T_{c}(4(2n-1)+1)$. 
As in the previous paragraph, a Trotter sequence yields a generic $U_{g}(\bold{a})\in \text{SU}(2^{2n})$ up to an error $\epsilon$ so that for the combined system an upper bound on $T(\bold{a})$ is obtained, where the explicit form is given in \cite{OnlineMaterial}. 
For $\Vert\bold{a}\Vert_{\infty}^{2}=\mathcal O(1)$ and $l(\bold{a})=\mathcal O(n)$, we conclude that $T(\bold{a})\leq \mathcal O(T_{c}(2n)^{5})$. It immediately follows that upon combining $L$ qubit graphs, each consisting of $n$ qubits, through $L-1$ controllable two body interactions, then the time $T(\bold{a})$ to implement $U_{g}(\bold{a})\in \text{SU}(2^{Ln})$ scales at most as $\mathcal O(T_{c}(Ln)^{5})$. Thus, as a sufficient criterion for a qubit system being scalable we have the following result. \\

\emph{
Using $L-1$ controllable two body interactions every $U_{g}(\bold{a})\in \text{SU}(2^{Ln})$ can be implemented on a $Ln$-qubit network in a time which scales at most polynomially in $L$ if $\Vert\bold{a}\Vert_{\infty},~l(\bold{a}) \leq \mathcal O(\text{poly}(Ln))$} . \\

 Concatenating blocks of qubits through controllable two body interactions allows for scaling the total system so that the gate set $\mathcal R_{\text{poly}}$ can be implemented efficiently on the combined system. This situation emphasizes the importance of being able to control two body interactions. However, an allied question is whether a qubit graph exists for which a few local controls are sufficient to implement $\mathcal R_{\text{poly}}$ efficiently. To address this goal requires identifying a system and a number of controls that allow for implementing each two qubit unitary \eqref{eq:twobodyunitary} in a time that scales at most polynomially in the number of qubits. Based on a decoupling scheme, for a $n$- qubit system in the previous paragraph, this goal is always possible using $2n$ controls. Now we show that for a star shaped graph the number of controls can be reduced to $n+1$.    

\emph{Reducing the number of controls --} Consider a star shaped graph described by the drift Hamiltonian 
$H_{0}=J\sum_{i=2}^{N+1}(\sigma_{x}^{(1)}\sigma_{x}^{(i)}+\sigma_{y}^{(1)}\sigma_{y}^{(i)})+J\sum_{i=2}^{N+1}\sigma_{y}^{(i)}$,	 where for the sake of simplicity the couplings and the energy splittings are assumed to be all given by $J$. Control is exerted through $\{\sigma_{x}^{(1)},~\sigma_{y}^{(1)},~\sigma_{z}^{(i)}\}$,~$i=2,\cdots, N+1$.  For an illustration of such a graph we refer to the online material \cite{OnlineMaterial}. Through decoupling using a string of $\sigma_{z}^{(i)}$ we can instantaneously implement unitaries corresponding two body interaction terms $H_{k}=(\sigma_{x}^{(1)}\sigma_{x}^{(k)}+\sigma_{y}^{(1)}\sigma_{y}^{(k)}+\sigma_{y}^{(k)})$. Further decoupling with $\sigma_{x}^{(1)}$, $\sigma_{z}^{(k)}$ and instantaneous local rotations of qubit $1$ and qubit $k$ yield unitaries corresponding to $\sigma_{\alpha}^{(1)}\sigma_{x}^{(k)},~\sigma_{\beta}^{(1)}\sigma_{y}^{(k)}$ and $\sigma_{y}^{(k)}$. Recall that  $\Delta t=\frac{\pi}{2J}$ units of time are needed to create a unitary operation generated by $[\sigma_{\alpha}^{(1)}\sigma_{x}^{(k)},\sigma_{y}^{(k)}]$. Further note that a unitary operation $U_{g}=\exp(a_{i}S_{i})$ with $S_{i}\in\{i\sigma_{\alpha}^{(1)}\sigma_{\beta}^{(k)}\}$ takes at most $T_{c}=\frac{1}{J}(\frac{\pi}{2}+|a_{i}|)$ time, where $\{i\sigma_{\alpha}^{(1)}\sigma_{\beta}^{(k)}\}$ forms a generating set. In order to obtain a unitary operation corresponding to a commutator $[S_{1},S_{2}]$ requires $\Delta t\leq \frac{3\pi}{J}$ units time.  Thus, the time $\tau(a_{i}B_{i})$ to create $U_{g}=\exp(a_{i}B_{i})$ with $B_{i}\in \mathfrak{su}(2^{N+1})$ is upper bounded by $\tau(a_{i}B_{i})\leq D(B_{i})\frac{3\pi}{J}+\frac{1}{J}(\frac{\pi}{2}+|a_{i}|)$. By upper bounding the depth we then find
$	\tau(a_{i}B_{i})\leq \frac{1}{J}\left(\frac{\pi}{2}(12(n-2)+1)+|a_{i}|  \right)$,
 where $n=N+1$ is the total number of qubits. Therefore, every $U_{g}=\exp(a_{i}B_{i})$ can be implemented in a time that scales at most linearly in the number of qubits. Again using a Trotter sequence finally permits concluding that for the star shaped graph every $U\in \mathcal R_{\text{poly}}$ can be implemented efficiently up to some error $\epsilon$ using only $n+1$ controls. The star shaped graph model is of particular importance since it is used to describe the interaction of an electron spin in a nitrogen vacancy center with the surrounding nuclear spins \cite{NVmodel, NVmodel2, NVQuantum1, NVQuantumC2}, and, in general, quantum dots in a spin bath \cite{QubitSpinBath1, QubitSpinBath2, QubitSpinBath3}.

\emph{Conclusions --}
We have characterized the set of gates that (provably) can be implemented in polynomial time on a generic qubit network where each qubit is controlled locally using time dependent fields. The characterization relied on the assumption that the control fields are unconstrained in strength. Further investigations regarding the importance of this assumption, as well as an assessment of the tightness of the derived bound can be found in the online material \cite{OnlineMaterial}. The results directly apply to state preparation and quantum simulation.   
The control of two body interactions allows for concatenating blocks of qubits so that the total system can be controlled efficiently, thereby paving the way towards scalable quantum devices. Moreover, we have identified a model, with applications in nitrogen vacancy centers, for which the efficiently implementable gate set can be realized with fewer local controls. 
 Future research is required for characterizing qubit graphs and controls permitting the efficient implementation of $\mathcal R_{\text{poly}}$. An interesting goal would be the determination of the minimum number of controls required and the corresponding graph topologies.

\newpage

{\em Acknowledgements.} -- C. A. acknowledges the NSF (grant CHE-1464569) and fruitful discussion with Daniel Burgarth and Benjamin Russell. H. R. acknowledges the ARO (grant W911NF-16-1-0014).


\bibliographystyle{apsrev}

\onecolumngrid
\section*{Supplemental material}
\label{ref:der}

In this supplemental material we provide a more detailed derivation of Eq. (1), (6) and (8) in the main paper, a comparison of the developed bound with exact results for the minimum gate time found for low dimensional spin systems in \cite{ExactCalc1,ExactCalc2}, and an illustration of a star-shaped graph for which the number of controls can be reduced while still being able to implement $\mathcal R_{\text{poly}}$. Using numerical gate optimization \cite{Grape, QuTip1, QuTip2} the tightness of the bound is additionally studied for higher dimensional spin systems for which the minimum gate time is not known. Smooth pulses with a finite amplitude are found that implement a desired gate with a gate time below the derived bound.    
\subsection*{Detailed derivation of Eq's (1), (6) and (8)}
We recall that through a decoupling sequence a generic two body interaction term $\sigma_{\alpha}^{(i)}\sigma_{\beta}^{(j)}$ can be selected instantaneously \cite{DecLloyd}, so that at time $t$ a unitary evolution $U(t)=\exp(-itg_{\alpha,\beta}^{(i,j)}\sigma_{\alpha}^{(i)}\sigma_{\beta}^{(j)})$ is obtained. It therefore takes $k/g_{\alpha,\beta}^{(i,j)}$ amount of time to generate a unitary operator 
\begin{align}
U_{\alpha,\beta}^{(i,j)}(k)=e^{\pm ik \sigma_{\alpha}^{(i)}\sigma_{\beta}^{(j)}},~~~~k\in \mathbb R_{+},~~~~\alpha,\beta\in\{x,y,z\}.
\end{align}
Through the sequence 
   \begin{align}
 \label{eq:sequenceapp}
 &U_{x,z}^{(2,3)\dagger}\left(k_{1}\right)U_{z,y}^{(1,2)}(k)U_{x,z}^{(2,3)}\left(k_{1}\right)\nonumber\\
 &=\exp(-ik[\cos(2k_{1})\sigma_{z}^{(1)}\sigma_{y}^{(2)}-\sin(2k_{1})\sigma_{z}^{(1)}\sigma_{z}^{(2)}\sigma_{z}^{(3)}]),
 \end{align}
the unitary operator $U=\exp(ik\sigma_{z}^{(1)}\sigma_{z}^{(2)} \sigma_{z}^{(3)})$ from the main text (Eq. (6)) is obtained for $k_{1}=\frac{\pi}{4}$, i.e., by a commutator of depth $1$.  Thus, it takes $k/{g_{z,y}^{(1,2)}}+\Delta t=k/{g_{z,y}^{(1,2)}}+\pi/(2g_{x,z}^{(2,3)})$ to create such a unitary operator. Introducing the smallest coupling constant as $J\equiv \min_{i,j,\alpha,\beta}\{g_{\alpha,\beta}^{(i,j)}\}$ shows that $\Delta t$ is generally upper bounded by $\frac{\pi}{2J}$. From the form of the sequence \eqref{eq:sequenceapp} it inductively follows that a nested commutator of depth $D(B_{i})$ takes $D(B_{i})\Delta t$ units of time. Thus,  the time $\tau(a_{i}B_{i})$ to implement a unitary operator $U=\exp(a_{i}B_{i})$ is upper bounded by 
\begin{align}
\label{eq:taBapp}
\tau(a_{i}B_{i})\leq \frac{1}{J}\left(D(B_{i})\frac{\pi}{2}+|a_{i}|\right), 	
\end{align}
where the $|a_{i}|/J$ term is an upper bound for the time $k/g_{\alpha,\beta}$ it takes to implement any unitary of the form Eq. (1), which is the starting point of the construction Eq. (2). 

  \begin{figure}
  \includegraphics[width=0.35\columnwidth]{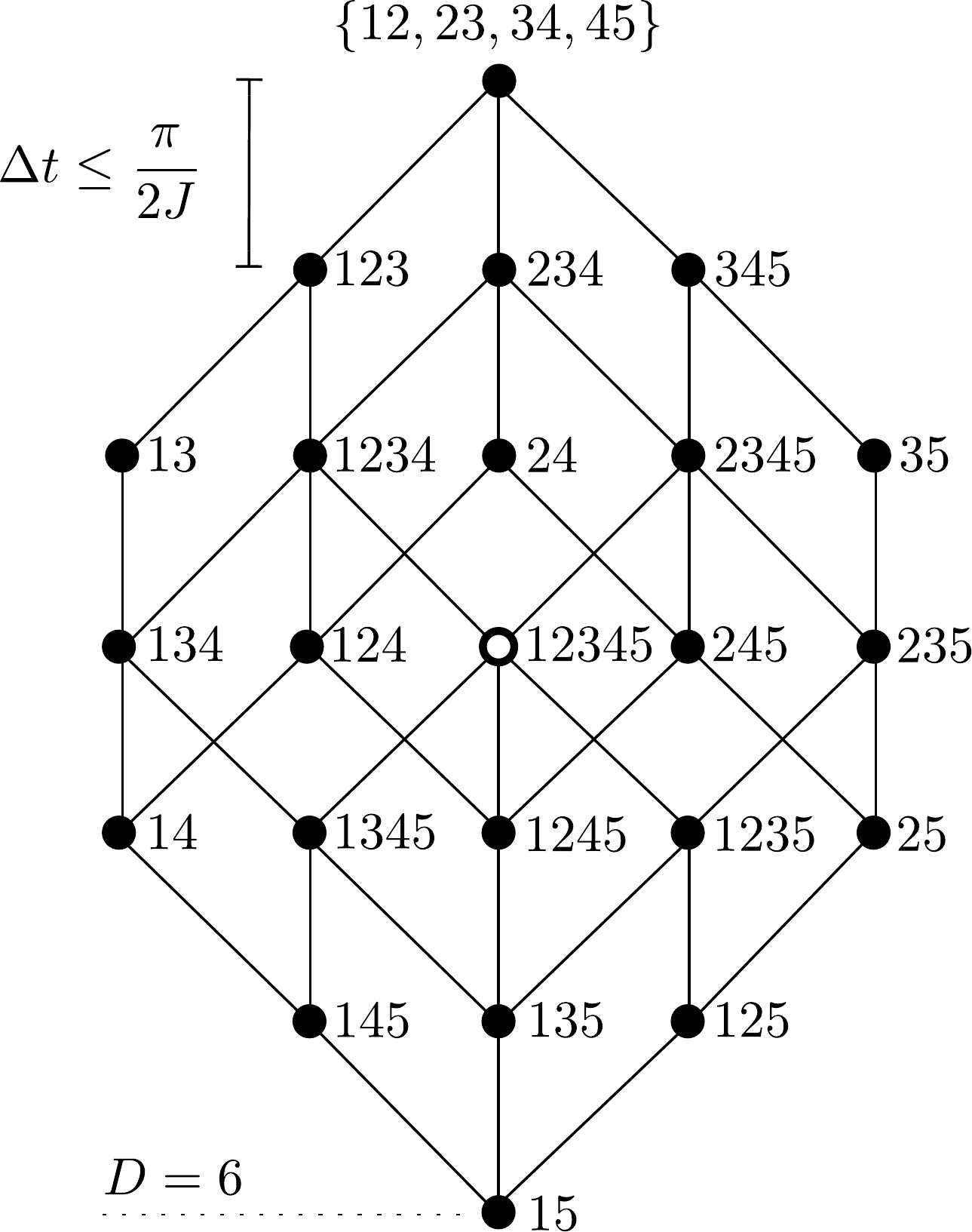}
 \caption{\label{fig:Lietree} Lie-tree diagram generated by two body interaction terms $\{\sigma_{\alpha}^{(i)}\sigma_{\beta}^{(i+1)}\}~,\alpha,\beta\in\{x,y,z\},~1\leq i\leq 4,$ for a 5 qubit system. From top to bottom, the vertices represent basis operators, which are strings of Pauli operators (up to local rotations), i.e. $\sigma_{\alpha}^{(1)}\sigma_{\beta}^{(2)}\cdots$, and that are obtained by a nested commutator of increasing depth. The numbers indicate on which qubit the operators act non-trivially and the edges represent the commutator of a Pauli string with a particular (two-body) Pauli string from the top vertex (the generating set) that increase or decreases the length of the Pauli string. Pauli strings that consist of a single Pauli operator and Pauli strings that were already obtained at a lower depth are not represented. Creating a commutator takes at most $\frac{\pi}{2J}$ units of time and every basis operator can be created through a nested commutator of at most depth $D=6$. The open circle vertex represents a 5 body interaction term (a Pauli string of length 5) from which all other basis elements can be created with a nested commutator of at most depth 3.}
\end{figure}

Before turning to bounding the depth $D(B_{i})$, we first upper bound the minimum gate time $T(\bold{a})$ for a generic $U_{g}\in \text{SU}(2^{n})$. This result will upper bound the time it takes to create unitary operators containing linear combinations of the $B_{i}$s. Along the  lines of Trotter, a unitary operation $U_{g}=e^{\Theta}$ can be created by applying the sequence $\mathcal G=\prod_{i=1}^{l(\bold{a})}e^{\frac{a_{i}}{m}B_{i}}$ $m$-times. For a $2^{n}$ dimensional system the (normalized) error $\epsilon=\frac{1}{\sqrt{2^{n+1}}}\Vert\mathcal G^{m}-U_{g} \Vert$, with $\Vert\cdot\Vert$ being the Hilbert-Schmidt norm for implementing $U_{g}$, is upper bounded by \cite{Trotter}, 
\begin{align}
\label{eq:Trotter}
\epsilon\leq \frac{1}{2m\sqrt{2^{n+1}}}\sum_{j>k}|a_{j}a_{k}|\Vert[B_{j},B_{k}]\Vert,	
\end{align}
where we assume that $\Theta$ contains at least two non-commuting basis operators. 
By introducing $\Vert\bold{a}\Vert_{\infty}= \max_{i}\{|a_{i}|\}$, we can further upper bound the right hand side, finding $\epsilon\leq \frac{l(\bold{a})(l(\bold{a})-1)\Vert\bold{a}\Vert_{\infty}^{2}}{2\sqrt{2}m}$. From \eqref{eq:taBapp} we have that the time $T(\mathcal G)$ to create $\mathcal G$ is upper bounded by $T(\mathcal G)\leq \frac{1}{J}\left(\frac{\Vert{\bold{a}}\Vert_{1}}{m}+\frac{\pi}{2}\sum_{i=1}^{l(\bold{a})}D(B_{i})\right)$,  where $\Vert\bold{a}\Vert_{1}=\sum_{i}|a_{i}|$ is the vector 1-norm. Consequently, the time to implement $U_{g}$ up to error $\epsilon$ is upper bounded by $T(\bold{a})\leq \frac{1}{J} \left(\Vert\bold{a}\Vert_{1}+m\frac{\pi}{2}\sum_{i=1}^{l(\bold{a})}D(B_{i})\right)$. With \eqref{eq:Trotter} we finally conclude that the minimum gate time $T(\bold{a})$ to implement $U_{g}=\exp(\sum_{i=1}^{l(\bold{a})}a_{i}B_{i})$ up to an error $\epsilon$ in a $n$ qubit system defined by Eq. (1) and Eq. (2) in the paper is upper bounded by  
\begin{align}
T(\bold{a})\leq \frac{1}{J}\left(\Vert\bold{a}\Vert_{1}+\frac{\pi}{2}\frac{\left(\sum_{j>k}|a_{j}a_{k}|\Vert[B_{j},B_{k}]\Vert\right)\left(\sum_{i=1}^{l(\bold{a})}D(B_{i})\right)}{2\sqrt{2^{n+1}}\epsilon}  \right),
\end{align}
which is the upper bound in Eq. (8) of the main text. Consider now the upper bound of depth $D(B_{i})$ for a system consisting of $n$ qubits. As schematically represented for $n=5$ qubits in the Lie tree diagram \cite{Elliot, Carenzspinstar} in Fig. 1, the sequence \eqref{eq:sequenceapp} leads to identifying that it takes a nested commutator of depth $(n-2)$ to create a basis operator that contains $n$ Pauli operators $\sigma_{\alpha}^{(1)}\sigma_{\beta}^{(2)}\cdots\sigma_{\delta}^{(n)}$ (open circle in Fig. \ref{fig:Lietree}). From this operator it takes another $(n-2)$ commutators to create any $B_{i}$, excluding the basis operators that correspond to a single Pauli operator. Thus, the depth is upper bounded by $D(B_{i})\leq 2(n-2)$ and since 
\begin{align}
\sum_{j>k}|a_{j}a_{k}|\Vert[B_{j},B_{k}]\Vert&\leq \frac{l(\bold{a})(l(\bold{a})-1)}{2}\max_{j,k}|a_{j}a_{k}|\Vert[B_{j},B_{k}]\Vert\nonumber \\
&\leq l(\bold{a})(l(\bold{a})-1)\max_{j,k}|a_{j}a_{k}|\Vert B_{j}B_{k}\Vert \nonumber \\
&\leq l(\bold{a})(l(\bold{a})-1)\Vert\bold{a}\Vert_{\infty}^{2}\sqrt{2^{n}}
\end{align}
we arrive at the upper bound Eq. (1) given in the introduction of the main body of the paper, i.e.,
\begin{align}
\label{eq:boundfinalapp}
T(\bold{a})\leq\frac{l(\bold{a})}{J}\left(\Vert \bold{a}\Vert_{\infty}+\frac{\pi l(\bold{a})(l(\bold{a})-1)(n-2)\Vert \bold{a}\Vert_{\infty}^{2}}{2\sqrt{2}\epsilon}   \right).	
\end{align}
For the concatenation of two qubit graphs through a controllable two body interaction, since here $\tau(a_{i}B_{i})\leq T_{c}(4(2n-1)+1)$ we analogously find 
\begin{align}
T(\bold{a})\leq T_{c}\frac{[4(2n-1)+1]l^{3}(\bold{a})[l(\bold{a})-1]\Vert\bold{a}\Vert_{\infty}^{2}}{2\sqrt{2}\epsilon}.
\end{align}

\subsection*{Comparison with known results and simulations to the bound tightness}
In this section we compare the derived upper bound with the exact minimum gate time obtained for a 3-spin Ising chain in \cite{ExactCalc2} and study the tightness of the bound for a 4-spin Heisenberg chain using numerical gate optimization \cite{Grape, QuTip1, QuTip2}. 
The drift Hamiltonian describing an $n$-spin Ising chain with nearest neighbour interactions reads 
\begin{align}
\label{eq:Ising}
H_{0}=\frac{\pi}{2} J\sum_{k=1}^{n-1}\sigma_{z}^{(k)}\sigma_{z}^{(k+1)},
\end{align}
where control is exerted through $\{\sigma_{x}^{(k)},\sigma_{y}^{(k)}\}_{k=1}^{n}$. For $n=3$ it was shown in \cite{ExactCalc2} that the minimum gate time $T$ to implement the goal operation $U_{g}=\exp(-i\kappa\frac{\pi}{4}\sigma_{z}^{(1)}\sigma_{z}^{(2)}\sigma_{z}^{(3)})$  is given by 
\begin{align}
\label{eq:exact}
T=\frac{\sqrt{\kappa(4-\kappa)}}{2J}.
\end{align}

  \begin{figure}[!h]
  \includegraphics[width=0.7\columnwidth]{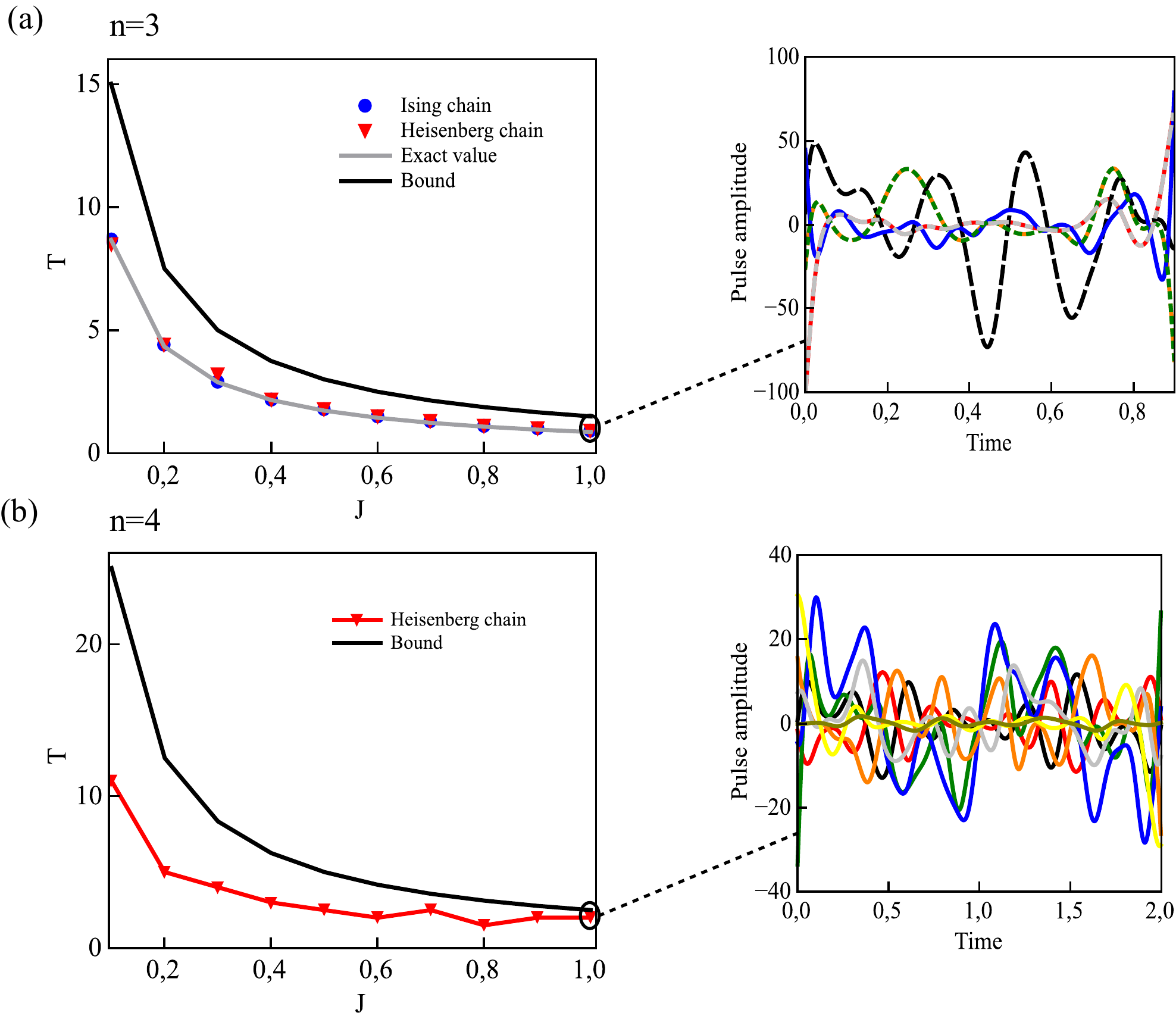}
 \caption{Minimum gate time $T$ as a function of the coupling strength $J$ and control pulses that implement the target gate for the time $T=0.9$ (a) and $T=2$ (b) and the coupling strength $J=1$. (a) For a 3-spin chain comparison of the bound \eqref{eq:Nbodybound} (black solid line) with the exact result \eqref{eq:exact} (grey solid line) and data obtained from numerical gate optimization for Ising interaction (blue circles) and Heisenberg interaction (red triangles). (b) For a 4-spin Heisenberg chain where the solid black line shows the bound \eqref{eq:Nbodybound}, noting that the exact value is not known, and the red triangles show the data obtained from numerical gate optimization.}
\end{figure}

In comparison, the bound Eq. (7) from the main body of the paper yields for implementing a unitary $U_{g}=\exp(-i\frac{\pi}{4}\kappa \sigma_{\alpha}^{(1)}\cdots \sigma_{\beta}^{(n)})$ generated by an $n$-body interaction term the upper bound
\begin{align}
\label{eq:Nbodybound}
T\leq\frac{(2(n-2)+|\kappa|)\pi}{4\tilde{J}},
\end{align}
valid for a generic $n$-spin chain, where $\tilde{J}$ is the smallest coupling constant present in the chain. Thus, for the 3-spin Ising chain described by \eqref{eq:Ising} with $n=3$, for $\kappa=1$ the bound above $\frac{3}{2J}$ defers from the exact value $T=\frac{\sqrt{3}}{2J}$ by a factor of $\frac{1}{\sqrt{3}}$.  We remark here that, in contrast to the exact value, the bound \eqref{eq:Nbodybound} is independent of the type of interaction and the number of spins present in the chain, and thus, it is also valid for a nearest neighbour isotropic Heisenberg type of interaction described by $H_{0}=\frac{\pi}{2} J\sum_{k=1}^{n-1}(\sigma_{x}^{(k)}\sigma_{x}^{(k+1)}+\sigma_{y}^{(k)}\sigma_{y}^{(k+1)})$.

In Fig. 2 we compared the upper bound \eqref{eq:Nbodybound} with the exact value \eqref{eq:exact} and data obtained from numerical gate optimization (triangles and circles) using the GRAPE algorithm \cite{Grape} implemented in the open source control package in QuTip \cite{QuTip1, QuTip2}. We remark here that the obtained numerical values are itself only an upper bound since the convergence of the optimization algorithm depends on the initial guess pulse.  Additionally the control pulses are shown that implement the target gate with high fidelity for the times $T=0.9$ for (a) and $T=2$ for (b) with coupling strength $J=1$ (marked with a circle).

 Fig. 2 (a) shows the minimum gate time $T$ for implementing $U_{g}=\exp(-i\frac{\pi}{4}\sigma_{z}^{(1)}\sigma_{z}^{(2)}\sigma_{z}^{(3)})$ with an error $\epsilon<10^{-3}$ in a 3-qubit Ising/Heisenberg chain (blue circles/red triangles) as a function of the coupling strength $J$. The solid black line shows the upper bound \eqref{eq:Nbodybound} and the grey line shows the exact value \eqref{eq:exact}. We see that for the Ising chain as well as for the Heisenberg chain the implementation of the target gate at the minimum gate time is possible with high fidelity through smooth controls with a finite amplitude. The methods \cite{DecLloyd} used for deriving the upper bound in the main text imply that the minimum gate time for both types of interactions is the same, which explains why the gate can be implemented at the minimum gate time \eqref{eq:exact} for both models. In fact, using a decoupling sequence, all types of nearest neighbour interactions can be mapped in no time into an Ising type interaction. Fig. 2 (b) shows the minimum gate time $T$ for implementing $U_{g}=\exp(-i\frac{\pi}{4}\sigma_{z}^{(1)}\sigma_{z}^{(2)}\sigma_{z}^{(3)}\sigma_{z}^{(4)})$ with an error $\epsilon<10^{-3}$ as a function of the coupling strength $J$ in 4-qubit Heisenberg chain for which the exact value of the minimum gate is not known.
      
\subsection*{Illustration of a star-graph with less controls}
In this section we illustrate a star-shaped graph for which the number of time dependent controls can be reduced while still being able to implement the set of polynomial gates $\mathcal R_{\text{poly}}$ given by Eq. (2) in the main paper.

  \begin{figure}[h]
  \includegraphics[width=0.40\columnwidth]{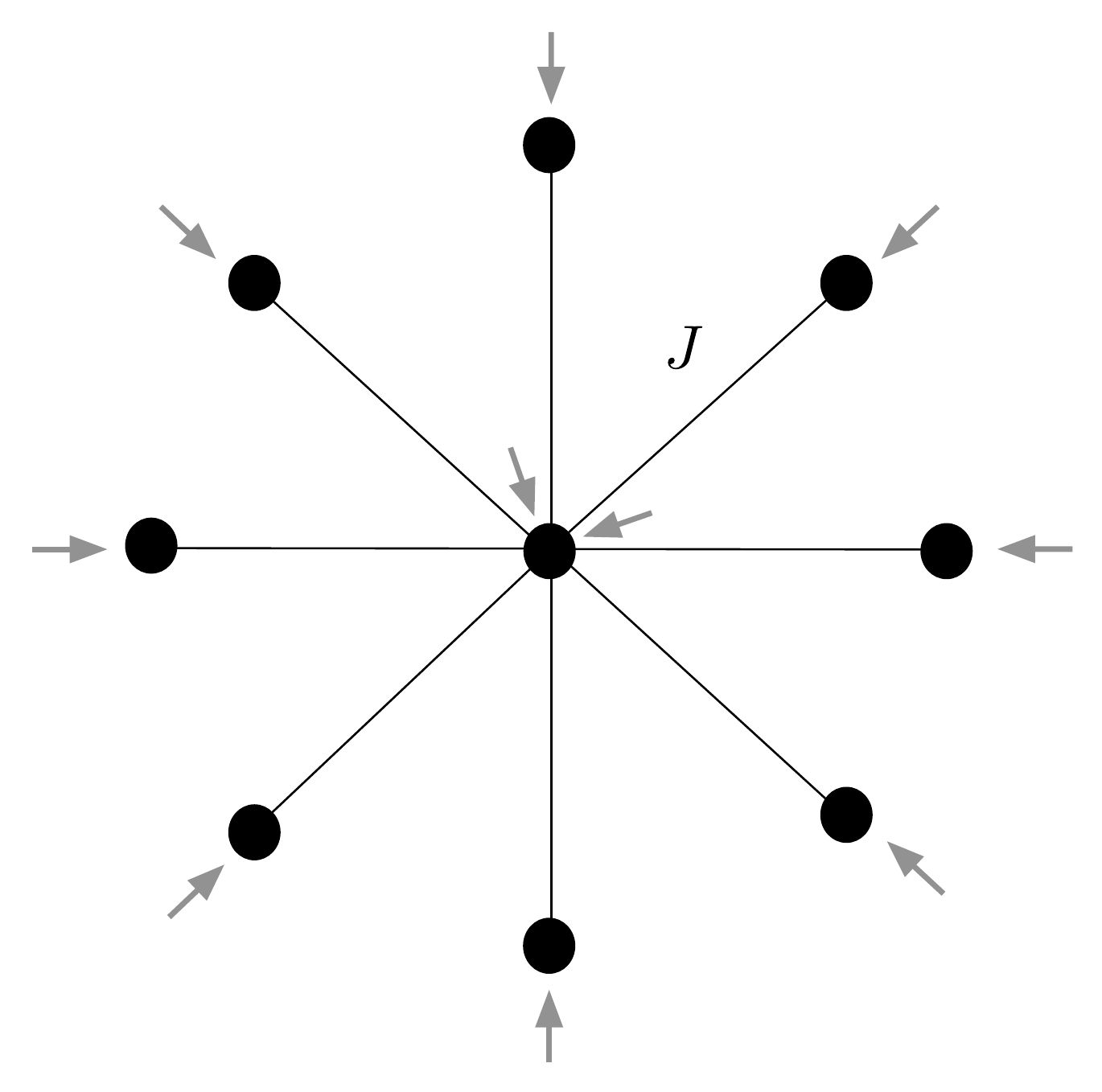}
 \caption{Illustration of a star-shaped graph consisting of $n$ spins (black circles) for which the number of controls (grey arrows) can be reduced from $2n$ to $n+1$ while the set of unitary gates $\mathcal R_{\text{poly}}$ can still be implemented in polynomial time. The central spin interacts with the surrounding spins through a isotropic Heisenberg interaction with interaction strength $J$ described by the Hamiltonian \eqref{eq:driftStar} and the controls are described by the Hamiltonian \eqref{eq:controlHStar}.}
\end{figure}

As illustrated in Fig. 3, we consider a central spin that interacts through a isotropic Heisenberg interaction with $N$ surrounding spins labeled by $i$. The surrounding spins have an energy splitting $J$ in $y$ direction so that the total Hamiltonian describing the system reads  
\begin{align}
\label{eq:driftStar}
H_{0}=J\sum_{i=2}^{N+1}(\sigma_{x}^{(1)}\sigma_{x}^{(i)}+\sigma_{y}^{(1)}\sigma_{y}^{(i)})+J\sum_{i=2}^{N+1}\sigma_{y}^{(i)}
\end{align}
The controls that allow for implementing $\mathcal R_{\text{poly}}$ are described by the control Hamiltonian 
\begin{align}
\label{eq:controlHStar}
H_{c}(t)=f_{1}(t)\sigma_{x}^{(1)}+h_{1}(t)\sigma_{y}^{(1)}+\sum_{i=2}^{N+1}f_{i}(t)\sigma_{z}^{(i)},
\end{align}
where $f(t)$ and $h(t)$ are the corresponding control fields.   

As shown in the main paper, the time $\tau(a_{i}B_{i})$ to implement a unitary operation generated by a basis operator $B_{i}$ scales linearly in $n$, i.e., $\tau(a_{i}B_{i})\leq \frac{1}{J}\left(\frac{\pi}{2}(12(n-2)+1)+|a_{i}|  \right)$. Thus, using a Trotter sequence we can concluded that every $U\in\mathcal R_{\text{poly}}$ can be implemented in a time that scales at most polynomial in the number of qubits.

\bibliographystyle{apsrev}

\begin{thebibliography}{99}
\bibitem{Comment1}
We remark here that this assumption, and thus the ability to instantaneously implement generic single qubit gates, is a reasonable approximation in the case where the strength of the control fields can be made much larger than the energy splittings/coupling constants of the systems being considered, which, for instance, is typically achieved in spin systems. In the online material we show for a 4-spin Heisenberg chain that smooth pulses with a finite amplitude can be obtained through numerical gate optimization with a gate time below the derived bound.
\end{thebibliography}

 \end{document}